\documentclass[aps,prd,twocolumn,superscriptaddress,showpacs,amsmath,amssymb,nofootinbib,preprintnumbers]{revtex4-1}
\usepackage{graphicx}
\usepackage{epstopdf}
\usepackage{float}
\usepackage{color}
\usepackage[toc,page]{appendix}
\usepackage[usenames,dvipsnames]{xcolor}
\usepackage{tensor}
\usepackage[normalem]{ulem}

\newcommand{\be}{\begin{equation}}
\newcommand{\ee}{\end{equation}}
\newcommand{\ba}{\begin{eqnarray}}
\newcommand{\ea}{\end{eqnarray}}

\newcommand{\beq}{\begin{equation}}
\newcommand{\eeq}{\end{equation}}
\newcommand{\beqa}{\begin{eqnarray}}
\newcommand{\eeqa}{\end{eqnarray}}

\begin{document}
\title{Lower-dimensional Gauss--Bonnet Gravity and BTZ Black Holes}

\author{Robie A. Hennigar}
\email{rhennigar@mun.ca}
\affiliation{Department of Mathematics and Statistics, Memorial University of Newfoundland, St. John's, Newfoundland and Labrador, A1C 5S7, Canada }

\author{David Kubiz\v n\'ak}
\email{dkubiznak@perimeterinstitute.ca}
\affiliation{Perimeter Institute, 31 Caroline St. N., Waterloo,
Ontario, N2L 2Y5, Canada}
\affiliation{Department of Physics and Astronomy, University of Waterloo,
Waterloo, Ontario, Canada, N2L 3G1}

\author{Robert B. Mann}
\email{rbmann@uwaterloo.ca}
\affiliation{Department of Physics and Astronomy, University of Waterloo,
Waterloo, Ontario, Canada, N2L 3G1}
\affiliation{Perimeter Institute, 31 Caroline St. N., Waterloo,
Ontario, N2L 2Y5, Canada}

\author{Christopher Pollack}
\email{cajpollack@edu.uwaterloo.ca}
\affiliation{Department of Physics and Astronomy, University of Waterloo,
Waterloo, Ontario, Canada, N2L 3G1}

\date{June 1, 2020}

\begin{abstract}
We consider the $D\to 3$ limit of Gauss--Bonnet gravity.  We find
two distinct but similar versions of the theory and obtain black hole solutions for each. For one
theory the solution is an interesting generalization of the BTZ black hole that does not have constant
curvature but whose thermodynamics is identical.  The other theory admits a solution that is asymptotically
AdS but does not approach the BTZ black hole in the limit of small Gauss--Bonnet coupling.  We also
discuss the distinction between our solutions and those obtained by taking a $D\to 3$ limit of solutions
to $D$-dimensional Einstein Gauss--Bonnet gravity.  We find that these latter  metrics are not solutions
of the theories we consider except for particular constraints on the parameters.
\end{abstract}

\maketitle

\section{Introduction}

Lovelock gravity~\cite{Lovelock:1971yv} is the most general theory of gravity built from the metric and Riemann curvature tensor that maintains second-order equations of motion for the metric. The theory consists of the cosmological and Einstein--Hilbert terms and introduces new corrections for each odd spacetime dimension above four. The first new such correction is the Gauss--Bonnet term
\be
{\cal G}=R_{abcd}R^{abcd}-4 R_{ab}R^{ab}+R^2\,,
\label{GBinv}
\ee
which is active in five or more dimensions. Importantly, these new contributions are either topological or identically zero for $D < 5$, singling out Einstein's theory as the most general second-order metric theory of gravity in four dimensions.

Recently there has been considerable interest generated by a proposal~\cite{Glavan:2019inb} of how to circumvent Lovelock's theorem. The idea is to treat the spacetime dimension as a parameter of the theory and rescale the Lovelock coupling constant according to
\be
(D-4)\alpha \to \alpha\,,
\ee
while taking the $D \to 4$ limit so as to obtain a nontrivial result in four dimensions. The idea as originally proposed suggested that solutions to the four-dimensional theory be constructed as limits of higher-dimensional solutions. In this way a number of enhanced symmetry $D=4$ metrics were obtained, each carrying an imprint of higher-curvature corrections inherited from their higher-dimensional counterparts. These include
 spherical black holes \cite{Glavan:2019inb,Kumar:2020uyz,Fernandes:2020rpa,Kumar:2020owy,Kumar:2020xvu},  cosmological solutions \cite{Glavan:2019inb, Li:2020tlo, Kobayashi:2020wqy}, star-like solutions  \cite{Doneva:2020ped}, radiating solutions \cite{Ghosh:2020vpc}, collapsing solutions \cite{Malafarina:2020pvl}
 all for Gauss--Bonnet gravity, with extensions to more higher-curvature Lovelock theories
 \cite{Casalino:2020kbt, Konoplya:2020qqh, Konoplya:2020der}. There are already a number of studies of the thermodynamic behaviour \cite{EslamPanah:2020hoj,Konoplya:2020cbv,Wei:2020poh,HosseiniMansoori:2020yfj,Zhang:2020qam,Singh:2020xju,Hegde:2020xlv}   and
physical properties \cite{Konoplya:2020bxa, Zhang:2020qew, Zhang:2020sjh, Wei:2020ght,Guo:2020zmf,Jin:2020emq,Heydari-Fard:2020sib,Liu:2020vkh,Islam:2020xmy, Roy:2020dyy, NaveenaKumara:2020rmi, Mishra:2020gce,Yang:2020czk, Ge:2020tid}
of these objects.

Subsequent work has called into question some aspects of this program~\cite{Gurses:2020ofy, Shu:2020cjw, Hennigar:2020lsl, Tian:2020nzb, Shu:2020cjw, Mahapatra:2020rds, Bonifacio:2020vbk, Arrechea:2020evj}. For example, in~\cite{Gurses:2020ofy} it was demonstrated that no purely geometric object exists that could serve as the field equations for the limiting theory,  while~\cite{Hennigar:2020lsl, Tian:2020nzb} focused on more complicated cosmological and Taub-NUT solutions, showing that the $D\to 4$ limit of solutions is not unique -- there are many ways by which a four-dimensional metric can be extended to higher dimensions and the limiting form of the solutions can retain information about the character of the extra dimensions.
Perhaps most convincing is the argument in \cite{Bonifacio:2020vbk} which shows that in four dimensions there are no other tree-level graviton scattering amplitudes than those of the Einstein gravity.
Taken in unison, these results suggest that the program as originally proposed is problematic.

However, these issues can be circumvented by considering more careful limits of the higher-dimensional theory itself. L{\"u} and Pang~\cite{Lu:2020iav} (see also \cite{Kobayashi:2020wqy})  used a Kaluza--Klein-like procedure to generate a four-dimensional limit of Gauss--Bonnet gravity by compactifying the higher-dimensional theory on a maximally symmetric space followed by taking the limit where the dimension of this space vanishes.  Subsequently~\cite{Fernandes:2020nbq, Hennigar:2020lsl} considered an alternative proposal -- a generalization of the technique used nearly 30 years ago by Mann and Ross to obtain a $D\to 2$ limit of general relativity~\cite{Mann:1992ar} -- to obtain a $D\to4$ limit of Gauss--Bonnet gravity. This approach has the advantage that it makes no assumptions regarding the character of the extra dimensions. Remarkably, the two approaches converge to the same limiting theory:\footnote{Strictly speaking, this equivalence holds only when the internal space used in the Kaluza-Klein reduction is flat. Reducing on more complicated internal spaces generates the following additional terms in the action:
\be\label{SDlam}
S_\lambda = \int d^D x \sqrt{-g} \Bigl(-2\lambda R e^{-2\phi}-12\lambda(\partial \phi)^2e^{-2\phi}-6\lambda^2 e^{-4\phi} \Bigr)\,,
\ee
where $\lambda$ is the curvature of the (maximally symmetric)  internal space.
}
\ba\label{SD}
S&=&\int d^D x \sqrt{-g}\Bigl[R-2\Lambda+\alpha\Bigl(\phi {\cal G}+4 G^{ab}\partial_a \phi \partial_b \phi\nonumber\\
&&\qquad\qquad\qquad -4(\partial \phi)^2 \Box \phi+2((\nabla\phi)^2)^2 \Bigr) \Bigr]\,.
\ea
The action~\eqref{SD} can reasonably be considered the closest thing to Gauss--Bonnet gravity that exists in four-dimensions. The theory possesses an additional scalar degree of freedom and is a special case of Horndeski theory~\cite{horndeski1974second} (see \cite{Schmidt:2018zmb} for a construction of conserved currents in these theories). The naive $D \to 4$ limit of the higher-dimensional spherically symmetric black hole solution to Gauss--Bonnet gravity is also a solution of this theory for a particular scalar configuration,\footnote{The same remains true in the presence of Maxwell field, adding
\be\label{Maxwell}
S_M=\int d^Dx \sqrt{-g}{\cal L}_M\,,\quad {\cal L}_M=-F=-F_{ab}F^{ab}
\ee
to \eqref{SD}, upon which one recovers the charged-GB black hole \cite{Fernandes:2020rpa}.}  though it is not the most general solution. The structure of more complicated solutions such as Taub-NUT are more subtle and do not coincide with the limits of higher-dimensional solutions~\cite{Hennigar:2020lsl}.

While much attention has been directed towards understanding the four-dimensional limit of Gauss--Bonnet gravity, there is no \textit{a priori} reason to exclude limits to lower dimensions.  The theory defined by~\eqref{SD} is not restricted to four dimensions, but can be studied also in lower dimensions. On the other hand, it is possible to consider alternatives to~\eqref{SD} obtained as limits of the Gauss--Bonnet term valid in three-dimensions or lower. It is our purpose here to study~\eqref{SD} and its possible alternatives in three dimensions. Gravity in lower dimensions has been a source of theoretical inspiration for many years due to its comparative simplicity. There is perhaps no better example of this than the well-known BTZ black hole~\cite{Banados:1992wn} which is a solution of Einstein's theory in three spacetime dimensions. The BTZ black hole has served as playground where many problems intractable in higher-dimensions can be solved~\cite{Ross:1992ba, Carlip:1995qv, Cardoso:2001hn, Konoplya:2004ik}, and there have been numerous studies of its generalizations in modified gravities~\cite{Townsend:2013ela, Shu:2014eza, Bravo-Gaete:2014haa,Chernicoff:2018hpb}. We start by considering to what extent the theory~\eqref{SD} admits generalizations of the BTZ black hole.

\section{Gauss--Bonnet BTZ black holes}

There is no logical obstruction to setting $D=3$ in the action~\eqref{SD} to obtain a $D=3$ version
of Gauss--Bonnet gravity.  In this section we consider this theory, noting that the quantity ${\cal G}$ identically vanishes for $D=3$.

\subsection{BTZ-like solutions}
It is known that all pure metric modified gravities admit the BTZ black hole as a solution~\cite{Gurses:2019wpb}.
Here we point out that the same is true in the theory~\eqref{SD}.
Consider the following BTZ ansatz:
\be\label{ds23d}
ds^2=-fdt^2+\frac{dr^2}{fh}+r^2\Bigl(d\varphi -\frac{J}{2r^2}dt\Bigr)^2\,,
\ee
where metric functions $f=f(r)$ and $h=h(r)$ and the solution is supported by the scalar field $\phi=\phi(r)$, and $J$ is a constant.
When inserted in the action \eqref{SD} (together with \eqref{SDlam}), one obtains an effective Lagrangian that has to be varied
w.r.t. $f, h, \phi$, yielding 3 equations of motion. In particular, focusing on the BTZ solutions that satisfy
the condition $h = 1$, the equation coming from $\delta f$ reads
\be\label{phiEq}
\alpha \Bigl[\phi'^2 J^2+4 r^3(\phi'^2+\phi'')(fr\phi'^2-f\phi'-\lambda r e^{-2\phi})\Bigr]=0\,.
\ee

 A simple solution of \eqref{phiEq} is obtained by setting $\phi=$const. Of course, for $\lambda=0$ one then recovers the standard BTZ black hole \cite{Banados:1992wn} with
\be\label{BTZlike}
f=-m+\frac{r^2}{\ell^2}+\frac{J^2}{4r^2}\,,\quad \Lambda=-\frac{1}{\ell^2}\,.
\ee
For $\lambda\neq 0$, the situation is slightly more complicated. Setting $\phi=0$ for simplicity we find that
\be
f=-m+\lambda r^2+\frac{J^2}{4r^2}
\ee
solves the remaining equations, provided the following constraint is satisfied:
\be\label{rotConstraint}
\lambda+\alpha\lambda^2+\Lambda=0\,.
\ee
This is just the constraint determining the embedding of maximally symmetric spaces in the theory when $\phi = 0$.

\subsection{Novel black holes}

Apart from the BTZ metric \eqref{ds23d}, there exist other types of black hole solutions to the equations of motion of \eqref{SD} with $D=3$ that have a non-trivial scalar field profile. To demonstrate this, we begin by considering the static case, $J=0$, and take $\lambda=0$. Eq. \eqref{phiEq} then simplifies to $(\phi'^2+\phi'')(r\phi'-1)=0$, and admits the following solution:
\be\label{profile}
\phi=\ln(r/l)\,,
\ee
where $l$ is an integration constant, upon which the spacetime will no longer be of constant curvature.
The equation  coming from $\delta \phi$ is then identically satisfied, while that from $\delta h$
yields
\be
2\alpha r ff'+r^3f'-2\alpha f^2+2\Lambda r^4=0\,,
\ee
which is integrable and has the solution
\be\label{f3d}
f_\pm=-\frac{r^2}{2\alpha}\left(1\pm \sqrt{1+\frac{4\alpha}{r^2}\left(\frac{r^2}{\ell^2}  - m\right)}\right)\,,
\ee
with $m$ a constant of integration.

Only the $f_-$ branch of the above admits a well-defined limit as $\alpha \to 0$:
\be
f_-=\frac{r^2}{\ell^2}-m -\frac{\alpha}{r^2}\Bigl( \frac{r^2}{\ell^2}-m\Bigr)^2+O(\alpha^2)\,,
\ee
and obviously reproduces the standard BTZ solution in the limit of small $\alpha$.
At large distances the metric behaves as
\be
f=\frac{r^2}{2\alpha}(K-1)
-\frac{m}{K}+O(1/r^2)\,,\quad
K=\sqrt{1+\frac{4\alpha}{\ell^2}}\,,
\ee
which yields  $\ell^2_{\tiny \rm eff}=2\alpha/(K-1)$ as the effective cosmological constant.

A well-defined black hole solution requires $\alpha > -\ell^2/4$ since otherwise the metric function terminates at some finite value of $r$. When $\alpha$ is negative and in the range $-\ell^2/4 < \alpha < 0$ the effective cosmological constant is smaller than the corresponding one in Einstein gravity, while it is larger for $\alpha > 0$.  The horizon is located at $r_+$ where
\be
m=\frac{r_+^2}{\ell^2}\,,
\ee
and so, remarkably, coincides with the horizon of a BTZ black hole of the same mass. When $\alpha > 0$ the metric exhibits a branch singularity inside the horizon, analogous to that in higher-dimensional Gauss--Bonnet black holes. The derivatives of the metric blow up at the branch singularity, and thus a curvature singularity occurs at this point.  For $-\ell^2/4 < \alpha < 0$ the metric extends to $r = 0$, behaving as
\be
f(r) \sim \frac{\sqrt{-\alpha m} r}{\alpha} - \frac{r^2}{2 \alpha} + {\cal O}(r^3) \, ,
\ee
which in turn yields a divergence in the Kretschmann scalar as $r \to 0$:
\be
R_{abcd}R^{abcd} \sim -\frac{2m}{\alpha r^2} + \cdots.
\ee
Thus, there exists a curvature singularity at the origin for this regime of parameter values.

Let us finally note that for the same profile of the scalar field $\phi$, \eqref{profile}, one can easily integrate the remaining
equations even when $\lambda\neq 0$, to recover the following BTZ-like solution:
\be
f=\lambda l^2+\frac{r^2}{\ell_{\tiny \rm eff}^2}\,,\quad \ell^2_{\tiny \rm eff}=\frac{-1\pm \sqrt{1-4\alpha \Lambda}}{2\Lambda}\,.
\ee
Unfortunately, we were not able to solve for the most general solution of \eqref{phiEq}, which would perhaps give a rotating generalization of the solution \eqref{f3d}.

\subsection{Thermodynamics}
To compute the entropy of these black holes we employ the Iyer-Wald method~\cite{Wald:1993nt, Iyer:1994ys}. While problems in the application of the Wald method to Horndeski black holes were identified in~\cite{Feng:2015oea, Feng:2015wvb} (see also \cite{Peng:2015yjx}), those issues were attributed to the divergence of the scalar on the black hole horizon. In the cases we consider here no such pathology occurs, and the familiar Wald method can be applied.

To compute the Wald entropy, we first note that
\begin{align}
&16 \pi  P_{ab}^{cd} \equiv \frac{\partial {\cal L}}{\partial R_{cd}^{ab}} \\
&= \left[1 - 2 \alpha (\partial \phi)^2 - 2 \lambda \alpha e^{-2 \phi} \right] \delta_{[a}^{[c} \delta_{b]}^{d]}
+ 4 \alpha \delta_{[a}^{[c} \nabla_{b]}\phi \nabla^{d]}\phi\,, \nonumber
\end{align}
where a contribution from the Gauss--Bonnet term is absent since ${\cal G}$ vanishes identically in three dimensions. Note also that here we have introduced a factor of $16 \pi$, putting Einstein--Hilbert term in canonical form so that the Wald entropy associated to it will be $A/4$.

The Wald entropy is obtained by integrating $P_{ab}^{cd}$ over the bifurcation surface of the black hole:
\be
S = -2 \pi \int_{\mathcal{H}}  d^{D-2} x \sqrt{\gamma} \left[ P_{ab}^{cd} \hat{\epsilon}^{ab} \hat{\epsilon}_{cd}  \right] \, ,
\ee
where $\hat{\epsilon}_{ab}$ is the binormal to the horizon normalized so that $\hat{\epsilon}_{ab} \hat{\epsilon}^{ab} = -2$. In the present case we have $\hat{\epsilon}_{ab} = 2 t_{[a} r_{b]}$ where $t_{a}$ and $r_{a}$ are the respective components of the one-forms $dt$ and $dr$.
A simple computation shows that $P_{ab}^{cd} \hat{\epsilon}^{ab} \hat{\epsilon}_{cd} = \frac{1}{4 } \left[1 - \frac{2 \alpha}{r^2} (\partial_\varphi \phi)^2 - 2 \alpha \lambda e^{-2 \phi} \right] $.  We thus find
(for scalars independent of   $\varphi$ )
\begin{align}\label{Swald}
S =&\, \frac{\pi r_+}{2} \Bigl[1 - 2 \lambda \alpha e^{-2 \phi} \Bigr] \, ,
\end{align}
where in this expression it is to be understood that the fields are evaluated on the horizon. Eq. \eqref{Swald}
 is valid  provided the scalar configuration is regular on the horizon.

Let us come first to the thermodynamics of the BTZ-like solution \eqref{BTZlike}. In the case $\lambda = 0$, the thermodynamics is identical to that of the familiar BTZ black hole in Einstein gravity:
\ba
M&=&\frac{m}{8}=\frac{r_+^2}{8\ell^2} + \frac{J^2}{32 r_+^2}\,,\quad T=\frac{f'}{4\pi}=\frac{r_+}{2\pi \ell^2} - \frac{J^2}{8 \pi r_+^3} \,,\quad \nonumber\\
S&=&\frac{\pi r_+}{2}\,, \quad P=\frac{1}{8\pi \ell^2}\,,\quad V=\pi r_+^2\,, \quad \Omega = \frac{J}{16 r_+^2}  \, .
\ea
We do not discuss in detail here the thermodynamics of the $\lambda \neq 0$, $\phi = 0$ solution
as this is purely the embedding of the usual BTZ black hole into the theory. Instead, we focus on the thermodynamics of the novel solution \eqref{f3d} constructed in the previous section.

Turning to the thermodynamics of  \eqref{f3d}, we find
\ba
M&=&\frac{m}{8}=\frac{r_+^2}{8\ell^2}\,,\quad T=\frac{f'}{4\pi}=\frac{r_+}{2\pi \ell^2}\,,\quad S=\frac{\pi r_+}{2}\,,\nonumber\\
P&=&\frac{1}{8\pi \ell^2}\,,\quad V=\pi r_+^2\,,\quad \psi_\alpha=0\,,
\ea
which is  identical to that of the familiar BTZ black hole in Einstein gravity, albeit with the additional potential $\psi_\alpha$.  This is quite intriguing -- even though the curvature is not constant, the thermodynamic parameters are the same for any value of $\alpha$. We also obtain
\be
\delta M=T\delta S+V\delta P+\psi_\alpha \delta \alpha\,,\quad 0=TS-2PV+2\psi_\alpha \alpha\,,
\ee
which are the standard Smarr and first law relations.


It is known that the thermodynamic properties of higher-dimensional Lovelock black branes are identical to those of black branes in Einstein gravity~\cite{Cadoni:2016hhd,Hennigar:2017umz}. The observations here are consistent with this property, albeit now extended to lower dimensions.

\subsection{Other solutions}

It is interesting to note that the solution \eqref{f3d} coincides (upon setting $\alpha\to -\alpha)$  with the $\kappa=0$  metric derived in \cite{Konoplya:2020ibi} by taking the $D\to 3$ limit of the higher-dimensional Gauss--Bonnet solution:
\be\label{otherSols}
f_{\epsilon_f} = \kappa-\frac{r^2}{2\alpha}\Bigl(1+\epsilon_{f}\sqrt{1-4\alpha\Lambda -\frac{4\alpha(\kappa-\Lambda r_+^2)}{r^2}+\frac{4\alpha^2\kappa^2}{r^2r_+^2}}\Bigr)\,,
\ee
where $\epsilon_f=\pm$ and $\kappa = -1, 0, +1$.  It is natural to probe whether the $\kappa=\pm 1$ metrics are also solutions of the theory \eqref{SD}.

To test this, we consider $\lambda\neq 0$ and expand the theory~\eqref{SD} to include \eqref{SDlam}. We find that demanding Eq.~\eqref{otherSols} solves the field equations forces
\be
\phi(r) = \frac{1}{2} \ln \left[C_1 r^2 + 2 C_2 \right] \, , \quad
\kappa=\frac{(1\pm\sqrt{1-4\alpha \Lambda})r_+^2}{2\alpha}\,,
\ee
where
\begin{align}
C_1 &= - \frac{2 \alpha \lambda}{\kappa r_+^2} \, ,
\quad C_2 = - \frac{ \epsilon_f \alpha \lambda}{ (1 + \sqrt{1-4 \alpha \Lambda})}  \, .
\end{align}
However, irrespective of the further constraints required to make $\kappa = \pm 1$, with these constraints above the metric reduces essentially to the BTZ geometry:
\be
f = - \lambda \Bigl(\frac{r^2}{2 C_2} + \frac{1}{C_1} \Bigr) \, .
\ee
Notwithstanding the parameter restrictions required to make this a black hole, note that the scalar diverges on the horizon and thus we cannot analyze the thermodynamics according to the usual Wald prescription.

The upshot of this is that the metric functions~\eqref{otherSols} with $\kappa = \pm 1$ are not solutions of the theory~\eqref{SD} with curvature corrections~\eqref{SDlam} except in the limiting cases where the couplings are constrained so that~\eqref{otherSols} reduces to the BTZ metric. We emphasize that this does not mean that those metrics are not solutions of \textit{some} lower-dimensional limit of Gauss--Bonnet gravity, but rather that they are not solutions to the (arguably) simplest theory~\eqref{SD}. In the remainder of this paper we consider alternate limits of Gauss--Bonnet gravity to lower dimensions and explore these limiting theories for simple solutions.

\section{Other $D\to 3$ limits of Gauss--Bonnet Gravity}

There are two methods by which the theory~\eqref{SD} can be obtained. One is via a dimensional reduction prescription where a $D$-dimensional theory is reduced on a $(D-p)$-dimensional maximally symmetric internal space followed by the limit $D \to p$~\cite{Lu:2020iav}. This leads to the action~\eqref{SD} with curvature correction~\eqref{SDlam}. Here $p$ refers to the dimensionality of the action after dimensional reduction. Thus, \eqref{SD} can be regarded as a $D \to 4$ limit of Gauss--Bonnet gravity (with $p = 4$) or a $D \to 3$ limit of Gauss--Bonnet gravity (with $p = 3$).

As shown recently \cite{Fernandes:2020nbq, Hennigar:2020lsl}, the $D=4$ Gauss--Bonnet gravity \eqref{SD} can be also obtained without dimensional reduction. The method is a generalization of one applied many years ago \cite{Mann:1992ar} to obtain a $D\to 2$ limit
of General Relativity. The essence of this prescription is as follows. One starts with the action --- or part of the action --- of interest and conformally transforms it. The transformed action is then expanded around the spacetime dimension of interest, and counterterms are added to the action to eliminate total derivative terms. The procedure concludes with a rescaling of the  couplings and the limit is taken to the spacetime dimension of interest. This procedure applied to Gauss--Bonnet gravity as $D \to 4$ yields the action $\eqref{SD}$~\cite{Fernandes:2020nbq, Hennigar:2020lsl}.

Here we apply the same approach to obtain a direct $D\to 3$ limit of Gauss--Bonnet gravity.  Consider
\be\label{SDGB}
S_D^{GB} = \alpha \left(\int d^Dx \sqrt{-\tilde g}\tilde{\cal G} - \int d^Dx \sqrt{-g}e^{\phi}{\cal G}\right)\,,
\ee
where the tilde quantities correspond to the conformally rescaled metric
\be
\tilde g_{ab}=e^{-2\phi} g_{ab}\,,
\ee
and the second term identically vanishes in $D<4$ dimensions. By expanding \eqref{SDGB} using formulae in App.~\ref{appA}, and rescaling the Gauss--Bonnet coupling
as
\be
(D-3)\alpha\to \alpha\,,
\ee
we find a finite action $S_3^{\cal{G}} = \lim_{D\rightarrow3} S_D^{GB}$, given by
\be\label{coget}
S_3^{\cal{G}}=\alpha\int d^3x \sqrt{-g}e^\phi\Bigl[-4 G^{ab}\partial_a \phi \partial_b \phi+2(\partial \phi)^2\Box \phi\Bigr]\,.
\ee
If we had instead applied the conformal trick to the full action, including also the Einstein--Hilbert term, then we would have arrived at the following result:
\be\label{S3c}
S_3^{(1)}=\int d^3x \sqrt{-g}e^{-\phi}\Bigl[R+2(\partial \phi)^2-2\Lambda e^{-2\phi}\Bigr] +S_3^{\cal{G}}\,.
\ee
If the conformal transformation is then undone  by setting $g_{ab}\to \exp(2\phi) g_{ab}$,  followed by the transformation $\phi \to - \phi $ and $\alpha \to - \alpha$ the action~\eqref{S3c} then reduces precisely back to~\eqref{SD}. This result is quite interesting as it suggests an element of universality to the lower dimensional limit of Gauss--Bonnet gravity.



A theory similar to \eqref{coget} can be obtained also via the Kaluza-Klein approach~\cite{Lu:2020iav}. This time, one considers a dimensional reduction on a $(D-p)$-dimensional flat space, rescales the Gauss--Bonnet coupling according to $(D-p -1) \alpha \to \alpha$, and takes the limit $D \to p + 1$. Thus, setting   $p = 3$, one
obtains\footnote{This theory is possibly supplemented by internal space curvature terms, see Eq. (11) in \cite{Lu:2020iav}.} \cite{Lu:2020iav}
\be\label{S3b}
S_3^{(2)}=\int d^3x \sqrt{-g}e^\phi(R-2\Lambda)+S_3^{\cal{G}}\, .
\ee

The Gauss--Bonnet portion of the action is the same in each case, but the treatment of the Einstein--Hilbert terms in actions  \eqref{S3b} and \eqref{S3c} differ between the two approaches. That is, the Einstein frame of~\eqref{S3b} does not coincide with the theory~\eqref{SD} but includes an additional kinetic term in the Einstein--Hilbert part of the action. Nonetheless, it is remarkable that the limiting forms of the Gauss--Bonnet density is identical between the two approaches. We emphasize the considerable difference between the two methods. From the perspective of the conformal trick~\eqref{coget} is the $D \to 3$ limit of Gauss--Bonnet gravity, while from the Kaluza-Klein perspective this is obtained as a $D \to 4$ limit of Gauss--Bonnet gravity dimensionally reduced on a  one-dimensional space.

We find that
the theory~\eqref{S3b} admits exotic black hole  solutions. For static solutions we obtain
\be
r^3(\phi''+\phi'^2)(2\alpha fr \phi'^2-4\alpha f \phi' +r )=0\,,
\ee
which replaces \eqref{phiEq}, while we do not present the remaining equations
here.  The full equations admit the following special solution:
\ba\label{fDto3}
f_\pm&=&\frac{r^2}{2\alpha}\Bigl(1\pm \sqrt{1+\frac{4}{3}\alpha \Lambda +\frac{4\alpha \zeta}{r^3}}\Bigr)\,,  \\
\phi&=&\ln(r/l)  \,,\nonumber\\
\ea
 where $\zeta$ and $l$ are constants of integration. This solution is qualitatively distinct from \eqref{f3d}.
 Indeed, the
  $f_-$ branch has a well-defined $\alpha\to 0$ limit,
\be
f_-=\frac{r^2}{3\ell^2}-\frac{\zeta}{r}+O(\alpha)\,,
\ee
but  does not approach the BTZ solution.  The metric function \eqref{fDto3} describes a black hole
whose horizon is located at $r_+$, where
\be
\zeta=\frac{r_+^3}{3\ell^2}\,.
\ee
The thermodynamics of this solution is complicated by a number of factors, including for example that it is a solution in the string frame rather than the Einstein frame. For these reasons we leave a full analysis of its properties for future work.

%

\section{Conclusions}

We have considered limits of Gauss--Bonnet gravity to three dimensions and constructed simple solutions to these theories that extend the familiar BTZ black hole of Einstein gravity.

The limits of Gauss--Bonnet gravity we have considered are obtained via two techniques. The first, used by L{\"u} and Pang~\cite{Lu:2020iav}, is a modification of Kaluza-Klein reduction. The second, used first by Mann and Ross~\cite{Mann:1992ar}, involves first conformally transforming the term of interest followed by coupling rescalings and the addition of counterterms to eliminate total derivatives. The final limits obtained for the Gauss--Bonnet term turns out to be identical in the two approaches (provided a flat internal space is used in the Kaluza-Klein case). In fact, the limit of this term is nothing more than a conformal transformation applied to the action~\eqref{SD}. However, the contribution of the Einstein--Hilbert term to the limiting theory differs between the two approaches and, moreover, the conceptual framework applied in each case is considerably different. Nonetheless, the fact that the two approaches yield the same result for the Gauss-Bonnet contribution is a point that we believe deserves further thought.

Let us note that an alternative ``holographic'' $D \to 3$ limit of the Gauss-Bonnet term as recently been studied in~\cite{Alkac:2020zhg}, leading to purely geometric three-dimensional theories. The connection between this approach and those considered here remains to be explored.

By exploring solutions of the limiting theories we have shown that the metrics obtained via the naive $D \to 3$ limit of higher-dimensional Gauss--Bonnet gravity are solutions to the limiting theory only in special cases. Namely, the naive limit of higher-dimensional black branes considered in~\cite{Konoplya:2020ibi} remain solutions, while the limit of black hole metrics with curved horizons are no longer solutions. We leave open the possibility that these metrics could be solutions to some modification of the theories we have considered here, or solutions with far more complicated (e.g. time dependent) scalar profiles, but it is not clear what those modifications would be or how they would be motivated.

This last observation raises important points for future consideration. A remarkable fact is that the naive $D \to 4$ limit of higher-dimensional static black hole metrics remain solutions to the theory~\eqref{SD} irrespective of their horizon topology. The fact that this is no longer generically true for $D = 3$ raises the question of what happens to higher-order Lovelock terms in four-dimensions. That is, is the problem we have observed particular to three dimensions, or is it a general feature for the limit of $n^{th}$-order Lovelock gravity in $D \le 2n - 1$ dimensions?

\section*{Acknowledgements}

This work was supported in part by the Natural Sciences and Engineering Research Council of Canada.
R.A.H.\ is supported by the Natural Sciences and Engineering Research Council of Canada
through the Banting Postdoctoral Fellowship program
D.K.\ acknowledges the Perimeter Institute for Theoretical Physics  for their support. Research at Perimeter Institute is supported in part by the Government of Canada through the Department of Innovation, Science and Economic Development Canada and by the Province of Ontario through the Ministry of Colleges and Universities.

\appendix

\section{Conformal transformations}\label{appA}

Consider the following conformal transformation in $D$ dimensions:
\be
\tilde{g}_{ab} = e^{\psi}g_{ab}\,.
\ee
Then we find the following transformations for the Riemann tensor and its contractions:
\begin{align}{\label{riemann}}
   \tensor{\tilde{R}}{_{abcd}} =&  e^{\psi}\Bigl(\tensor{R}{_{abcd}}+\frac{1}{4}\tensor{g}{_{bd}}\tensor{\psi}{_{;a}}\tensor{\psi}{_{;c}}- \frac{1}{4}\tensor{g}{_{ad}}\tensor{\psi}{_{;b}}\tensor{\psi}{_{;c}}- \frac{1}{2}\tensor{g}{_{bd}}\tensor{\psi}{_{;ac}}\nonumber\\
   &+\frac{1}{2}\tensor{g}{_{ad}}\tensor{\psi}{_{;bc}}- \frac{1}{4}\tensor{g}{_{bc}}\tensor{\psi}{_{;a}}\tensor{\psi}{_{;d}}+
   \frac{1}{4}\tensor{g}{_{ac}}\tensor{\psi}{_{;b}}\tensor{\psi}{_{;d}}+ \frac{1}{2}\tensor{g}{_{bc}}\tensor{\psi}{_{;ad}}
   \nonumber\\
   &-\frac{1}{2}\tensor{g}{_{ac}}\tensor{\psi}{_{;bd}}+ \frac{1}{4}\tensor{g}{_{ad}}\tensor{g}{_{bc}}\tensor{\psi}{_{;e}}\tensor{\psi}{_{;}^{e}}- \frac{1}{4}\tensor{g}{_{ac}}\tensor{g}{_{bd}}\tensor{\psi}{_{;e}}\tensor{\psi}{_{;}^{e}}\Bigr)\,,\\
\tilde{R}_{ab} =& R_{ab}+\frac{D-2}{4}\psi_{;a}\psi_{;b}-\frac{D-2}{4}g_{ab}(\partial \psi)^2\nonumber\\
    &-\frac{D-2}{2}\psi_{;ab}-\frac{1}{2}g_{ab}\Box \psi\,,\\
\tilde{R} =& e^{-\psi}\Bigl(R-\frac{(D-2)(D-1)}{4}(\partial \psi)^2-(D-1)\Box \psi\Bigr)\,,
\end{align}
and the Gauss--Bonnet invariant
\begin{align}{\label{gauss}}
    \tilde {\cal G} =
    &e^{-2\psi}\Bigl({\cal G}-\frac{1}{2}(D-4)(D-3)R(\partial \psi)^2-2(D-3)R_{ab}\tensor{\psi}{_;^a}\tensor{\psi}{_;^b}\nonumber\\
    &+\frac{1}{16}(D-4)(D-3)(D-2)(D-1)((\partial \psi)^2)^2\nonumber\\
    &-2(D-3)R\Box \psi+\frac{1}{2}(D-3)^2 (D-2)(\partial \psi)^2\Box \psi\nonumber\\
    &+(D-3)(D-2)(\Box \psi)^2+4(D-3)R_{ab}\tensor{\psi}{_;^{ab}}\nonumber\\
    &+(D-3)(D-2)\psi_{;a}\psi_{;b}\tensor{\psi}{_;^{ab}}-(D-3)(D-2)\psi_{;ab}\tensor{\psi}{_;^{ab}}
    \Bigr)\,.
\end{align}


\providecommand{\href}[2]{#2}\begingroup\raggedright\endgroup

\end{document}